\documentclass[12pt]{article}
\usepackage{amsfonts,amsmath,amsthm}
\usepackage{graphics}

\begin{document}
\title {Scaling laws for the movement of people between locations in a 
large city}

\author{G. Chowell$^{1,2}$, J. M. Hyman$^{2}$, S. Eubank$^{2}$, C. 
Castillo-Chavez$^{1,2}$\\ 
\footnotesize $^{1}$ Center for Nonlinear Studies (MS B258) \\
\footnotesize Los Alamos National Laboratory \\ 
\footnotesize Los Alamos, NM 87545, U.S.A. \\
\footnotesize $^{2}$ Department of Biological Statistics and
Computational Biology, Cornell University\\
\footnotesize Warren Hall, Ithaca, NY 14853, U.S.A
\thanks{Los Alamos Unclassified Report LA-UR-02-6658.} 
} 
\date{}
\maketitle

\begin{abstract} 
Large scale simulations of the movements of people in a ``virtual''
city and their analyses are used to generate new
insights into understanding the dynamic processes that depend on the 
interactions between people.  Models, based on these interactions, can
be used in optimizing traffic flow, slowing the spread
of infectious diseases or predicting the change in cell phone usage in 
a disaster. We analyzed cumulative and aggregated data generated from the simulated
movements of $1.6$ million individuals in a computer (pseudo agent-based)
model during a typical day in Portland, Oregon. This city is mapped into a 
graph with $181,206$ nodes representing physical locations such as 
buildings. Connecting edges model individual's flow between nodes. Edge weights are
constructed from the daily traffic of individuals moving between locations.
The number of edges leaving a node (out-degree), the edge 
weights (out-traffic), and the edge-weights per location (total
out-traffic) are fitted well by power law distributions. The power law
distributions also fit subgraphs based on
work, school, and social/recreational activities. The resulting
weighted graph is a ``small world'' and has scaling laws consistent 
with an underlying hierarchical structure. We also explore the time 
evolution of the largest connected component and the distribution of the component
sizes. We observe a strong linear correlation between the out-degree and total
out-traffic distributions and significant levels of clustering.
We discuss how these network features can be used to characterize social networks
and their relationship to dynamic processes.
\end{abstract} 

\section{Introduction}

Similar scaling laws and patterns have been detected in a great number
of systems found in nature, society, and technology. Networks of scientific 
collaboration \cite{Newman1}\cite{Newman2}\cite{BA2}, movie actors
\cite{WattsStrogatz}, cellular networks \cite{Jeong1}\cite{Jeong2}, food webs
\cite{Williams}, the Internet \cite{Faloutsos}, the \textit{World Wide Web}
\cite{Albert2,Kumar}, friendship networks \cite{Amaral1} and
networks of sexual relationships \cite{Liljeros} among others have
been analyzed up to some extent. Several common properties have been
identified in such systems. One such property is the short average
distance between nodes, that is, a small number of edges need to be
traversed in order to reach a node from any other node. Another
common property is high levels of clustering
\cite{WattsStrogatz,Strogatz}, a characteristic absent in random networks \cite{Bollobas}. 
Clustering measures the probability that the neighbors of a
node are also neighbors of each other. Networks with short average
distance between nodes and high levels of clustering have been dubbed 
``small worlds'' \cite{WattsStrogatz,Strogatz}. Power-law behavior in the
degree distribution is another common property in many real world
networks \cite{BA1}. That is, the probability that a randomly chosen
node has degree $k$ decays as $P(k) \sim k^{-\gamma}$ with $\gamma$ 
typically between $2$ and $3$. Barab\'{a}si and Albert (BA) introduced an algorithm capable of 
generating networks with a power-law connectivity distribution ($\gamma=3$). The BA 
algorithm generates networks where nodes connect, with higher probability, to 
nodes that have a accumulated higher number of connections and stochastically 
generates networks with a power-law connectivity distributions 
in the appropriate scale.\\

Social networks are often difficult to characterize because of
the different perceptions of what a link constitutes in the social
context and the lack of data for large social networks of more than a few
thousand individuals. Even though detailed data on the daily movement
of people in a large city does not exist, these systems have been
statistically sampled and the data used to build detailed simulations
for the full population. The insights gained by studying the simulated
movement of people in a virtual city can help guide research in
identifying what scaling laws or underlying structures may exist and
should be looked for in a real city. In this article we analyze a social mobility
network that can be defined accurately by the simulated movement 
of people between locations in a large city. We analyze the 
cumulative directed graph generated from the simulated movement of
$1.6$ million individuals \textit{in}
or \textit{out} of $181,206$ locations during a typical day in
Portland, OR. The $181,206$ nodes represent locations in the city and
the edges connections between nodes. The edges are weighted by daily
traffic (movement of individuals) \textit{in} or \textit{out} of these
locations. The statistical analysis of the cumulative network
reveals that it is a small world with power-law decay in the
out-degree distribution of locations (nodes). The resulting graph as well as
subgraphs based on different activity types exhibit scaling laws consistent with an
underlying hierarhical structure \cite{Erzsebet1, Erzsebet2}.
The out-traffic (weight of the full network) and the total out-traffic 
(total weight of the out edges per node) distributions are also fitted to power
laws. We show that the joint distribution of the out-degree and total out-traffic 
distributions decays linearly in an appropriate scale. We also explore
the time evolution of the largest component and the distribution of the
component sizes.

\subsection{Transportation Analysis Simulation System (TRANSIMS)}

\begin{figure}[h*]
  \begin{center}
    \scalebox{0.6}{\includegraphics{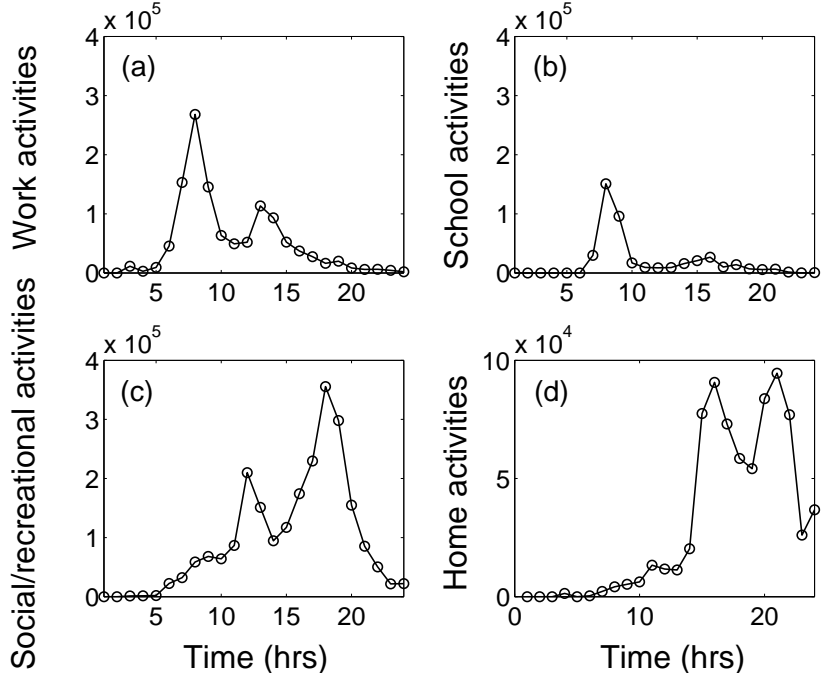}}
  \end{center}
 \caption{The number of people active in (a) work activities, (b) school activities, (d) social
   activities, and (d) home activities as a function of time (hours)
   during a `typical' day in Portland, Oregon.}
\label{myfig05}
\end {figure}

\indent TRANSIMS \cite{TRANSIMS} is an agent-based simulation model of
the daily movement of individuals in virtual region or city with a complete
representation of the population at the level of households and
individual travelers, daily activities of the individuals, and the
transportation infrastructure. The individuals are endowed with demographic characteristics taken
from census data and the households are geographically distributed
according to the population distribution. The
transportation network is a precise representation of the city's
transportation infrastructure. Individuals move across the
transportation network using multiple modes including car, transit,
truck, bike, walk, on a second-by-second basis. DMV records are
used to assign vehicles to the households so that the resulting
distribution of vehicle types matches the actual
distribution. Individual travelers are assigned a list of activities for the day
(including home, work, school, social/recreational, and shop
activities) obtained from the household travel activities survey for
the metropolitan area \cite{HSurvey} (Figure \ref{myfig05} shows the
frequency of four activity types in a typical day). Data on activities
also include origins, destinations, routes, timing, and forms of
transportation used. Activities for itinerant travelers such as bus drivers are 
generated from real origin/destination tables. \\
TRANSIMS consists of six major integrated modules: Population synthesizer, Activity
Generator, Router, Microsimulation and Emissions Estimator. Detailed
information on each of the modules is available
\cite{TRANSIMS}. TRANISMS has been designed to give transportation planners accurate,
complete information on traffic impacts, congestion, and
pollution.\\

\noindent For the case of the city of Portland, OR, TRANSIMS calculates the 
simulated movements of 1.6 million individuals in a typical day. 
The simulated Portland data set includes the time at which each 
individual leaves a location and the time of arrival to its 
next destination (node). These data are used to calculate the average
number of people at each location and the traffic between any two 
locations on a typical day. (Table $1$ shows a sample 
of a Portland activity file generated by TRANSIMS). Locations where
activities are carried out are estimated from observed land use
patterns, travel times and costs of transportation alternatives. These 
locations are fed into a routing algorithm that finds the minimum cost paths
that are consistent with individual choices \cite{Router-exp2, Router-theory, Router-exp1}. 
The simulation land resolution is of 7.5 meters. The simulator
provides an updated estimate of time-dependent travel times
for each edge in the network, including the effects of congestion, to
the \textit{Router and location estimation algorithms}
\cite{TRANSIMS}, which generate traveling plans. Since the entire
process estimates the demand on a 
transportation network from census data, land use data, and activity 
surveys, these estimates can thus be applied to assess the effects of hypothetical changes 
such as building new infrastructures or changing downtown parking prices.
Methods based on observed demand cannot handle such situations, since
they have no information on what generates the demand.
Simulated traffic patterns compare well to observed traffic and,
consequently, TRANSIMS provides a useful planning tool.\\

\noindent Until recently, it has been difficult to obtain useful
estimates on the structure of social networks. Certain classes of random graphs 
(scale-free networks \cite{BA1}, small-world networks \cite{Amaral1,Strogatz}, 
or Erdos-Renyi random graphs \cite{Bollobas,Erdos}), have been postulated as good representatives. 
In addition, data based models while useful are limited since they
have naturally focused on small scales \cite{Ackerman}. 
While most studies on the analysis of real networks are based on a single
snapshot of the system, TRANSIMS provides powerful time dependent data
of the evolution of a location-based network. 

\section{Portland's location-based network}
A ``typical'' realization by the Transportation Analysis 
Simulation System (TRANSIMS) simulates the dynamics of $1.6$ million 
individuals in the city of Portland as a directed network, where the 
nodes represent locations (i.e. buildings, households, schools, etc.) and 
the directed edges (between the nodes) represent the movement (traffic
due to activities) of individuals between locations (nodes). We have
analyzed the cumulative network of the whole day as well as cumulative
networks that comprise different time
intervals of the day. Here we use the term ``activity'' 
to denote the movement of an individual to the location
where the activity will be carried out. Traffic intensity is modeled
by the nonsymmetric mobility matrix $W = (w_{ij})$ of traffic weights
assigned to all directed edges in the
network ($w_{ij}=0$ means that there is no directed edge connecting 
node $i$ \textit{to} node $j$).

\begin{figure}[h*]
  \begin{center}
    \scalebox{0.7}{\includegraphics{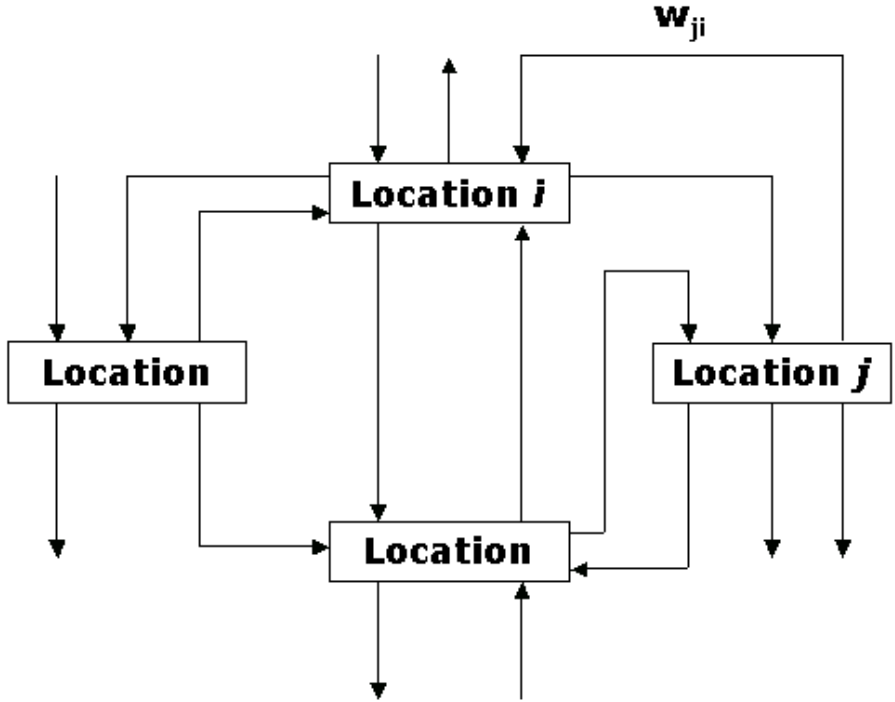}}
  \end{center}
 \caption{Structure of the location-based network of the city of
   Portland. The nodes represent locations connected via directed edges based on the traffic or 
   movement of individuals (activities) between the locations. The weights
   ($w_{ij}$) of the edges represent the daily traffic from location $i$ to location $j$.}
\label{myfig1}
\end {figure}

\begin{table}
\begin{center}
 \begin{tabular}{||c|c|c|c|c||}
 \hline
 Person ID & Location ID & Arrival time(hrs) & Departure time(hrs) & 
Activity type\\
 \hline 
 115 & 4225 & 0.0000 & 7.00 & home \\
 115 & 49296 & 8.00 & 11.00 & work \\
 115 & 21677 & 11.2 & 13.00 & work\\
 115 & 49296 & 13.2 & 17.00 & work \\
 115 & 4225 & 18.00 & 19.00 & home \\
 115 & 33005 & 19.25 & 21.00 & social/rec \\
 115 & 4225 & 21.3 &  7.00 & home \\
 220 & 8200 & 0.0000 & 8.50 & home \\
 220 & 10917 & 9.00 & 14.00 & school \\
 220 & 8200 & 14.5 & 18.00 & home \\
 220 & 3480 & 18.2 & 20.00 & social/rec \\
 220 & 8200 & 20.3 & 8.6 & home \\
 \hline 
 \end{tabular}
\end{center}
\label{table1}
\caption{Sample section of a TRANSIMS activity file. In this example, 
person $115$ arrives for a social recreational activity at location
$33005$ at $19.25$ o'clock and departs at $21.00$ o'clock.}
\end{table}
\section{Power law distributions}

\indent We calculate the statistical properties of a typical day in the location-based network 
of this vitual city from the cumulative mobility data generated by TRANSIMS (see
Table 2).\\
\indent The \textit{average out-degree} is $<k> =
\sum^{n}_{i=1} k_i/n$ where $k_i$ is the degree for 
node $i$ and $n$ is the total number of nodes in the network. For the
portland network $<k>=29.88$ and the \textit{out-degree
distribution} exhibits power law decay with scaling exponent 
($\gamma \approx 2.7$). The \textit{out-traffic} (edge weights) and the
\textit{total out-traffic} (edge-weights per node) distributions are
also fitted well by power laws. \\

\indent The \textit{average distance} between
nodes $L$ is defined as the median of the means $L_i$ of the shortest
path lengths connecting a vertex $i \in V(G)$ 
to all other vertices \cite{Watts}. For our network, $L=3.1$, which is small when compared to the
size of the network. In fact, the \textit{diameter} ($D$) of the graph (the largest of
all possible shortest paths between all the locations) is only $8$. $L$ and $D$
are measured using a breadth first search (BFS) algorithm
\cite{Sedgewick} ignoring the edge directions.\\

\indent  The \textit{clustering coefficient}, $C$, quantifies the
extent to which neighbors of a node are also neighbors of each other 
\cite{Watts}. The clustering coefficient of node $i$, $C_i$, is given by\\
$$C_i = \left | E(\Gamma_i) \right | / \binom{k_i} {2}$$\\
where  $\left | E(\Gamma_i) \right |$ is the number of edges in the 
neighborhood of $i$ (edges connecting the neighbors of $i$ not including 
$i$ itself) and $\binom{k_i} {2}$ is the maximal number of edges that 
could be drawn among the $k_i$ neighbors of node $i$. The clustering 
coefficient $C$ of the whole network is $C = \sum^{n}_{i=1}
C_i/n$. For a \textit{scale-free} random graph (BA model) \cite{BA1}
with $181,206$ nodes and $m=16$ \cite{comment1}, the clustering
coefficient $C_{rand} \approx \frac{(m-1)}{8}\frac{(ln N)^2}{N} \approx
0.0015$ \cite{Klemm1, Fronczak}. The clustering coefficient for our location-based network, ignoring
edge directions, is $C =0.0584$, which is roughly $39$ times larger
than $C_{rand}$. \\

\indent Highly clustered networks have been observed in
other systems \cite{WattsStrogatz} including the 
electric power grid of western US. This grid has a clustering coefficient 
$C=0.08$, about 160 times larger than the expected value for an equivalent 
random graph \cite{Watts}. The few degrees of separation between the
locations of the (highly clustered) network of the city of Portland
``make'' it a small world \cite{Strogatz,Amaral1,Watts}.\\

\begin{table}
\begin{center}
 \begin{tabular}{lc}
 Statistical properties & Value\\
 \hline
 Total nodes ($N$) & 181,206 \\
 Size of the cumulative largest component ($S$) & 181,192\\
 Total directed edges ($E$) & 5,416,005 \\
 Average out-degree ($<k>$) & 29.88 \\
 Clustering coefficient ($C$) & 0.0584 \\
 Average distance between nodes ($L$) & 3.1\\
 Diameter ($D$) & 8.0 
 \end{tabular}
\end{center}
\label{table2}
\caption{Statistical properties of the Portland's
location-based network. $S$ is the size of the largest component of the cumulative
network during the whole day.}
\end{table}

\begin{figure}[h*]
  \begin{center}
   \resizebox{6in}{3.5in}{\includegraphics{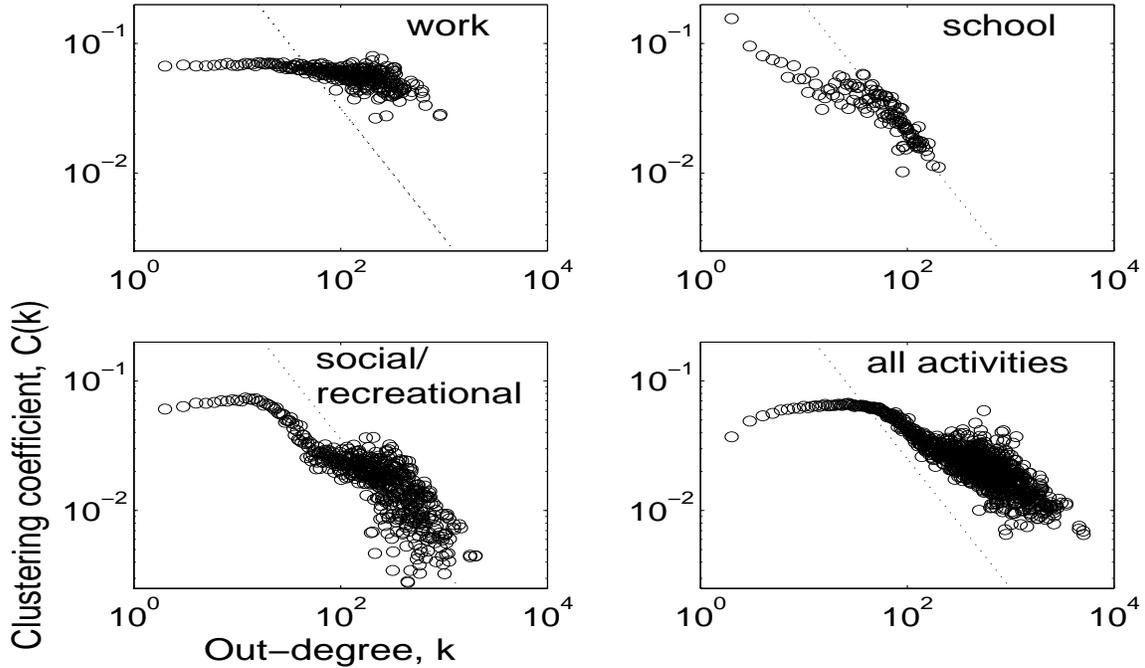}}
  \end{center}
 \caption{Log-log plots of the clustering coefficient as a function of 
the out-degree for subnetworks constructed from work 
activities, school activities, social activities, and all the 
activities. The dotted line has slope $-1$. Notice the scaling
$k^{-1}$ for the school and social/recreational activities. However,
for the subnetwork constructed from work activities, the clustering
coefficient is almost independent of the out-degree $k$.}
\label{myfig0}
\end{figure}

\indent Many real-world networks exhibit properties that are consistent 
with underlying hierarhical organizations. These networks have groups
of nodes that are highly interconnected with few or no edges connected
to nodes outside their 
group. Hierarchical structures of this type have been characterized by the 
clustering coefficient function $C(k)$, where $k$ is the node
degree. A network of movie actors, the semantic web, the \textit{World 
Wide Web}, the Internet (autonomous system level), and 
some metabolic networks \cite{Erzsebet1,Erzsebet2} have clustering 
coefficients that scale as $k^{-1}$. The clustering coefficient as a
function of degree (ignoring edge directions) in the Portland network
exhibits similar scaling at various levels of aggregation that
include, the whole network and subnetworks constructed by activity
type (work, school and social/recreational activities, see 
Figure \ref{myfig0}). We constructed subgraphs based
on activity types, that is, those subgraphs constructed from all the directed edges 
of a specific activity type (i.e work, school, social) during a typical
day in the city of Portland. 
The clustering coefficient of the subnetworks generated from work,
school, and social/recreational activities are: 
$0.0571$, $0.0557$, and $0.0575$, respectively. The largest clustering 
coefficient and closest to the overall clustering coefficient ($C = 
0.0584$) correponds to the subnetwork constructed from social/recreational 
activities. It seems that the whole network, as well as the selected activity 
subnetworks, support a hierarchical structure albeit the nature of such 
structure (if we choose to characterize by the power law exponent) is not universal. This agrees 
with relevant theory \cite{Erzsebet2}.\\

\begin{figure}[h*]
  \begin{center}
   \resizebox{5in}{3.5in}{\includegraphics{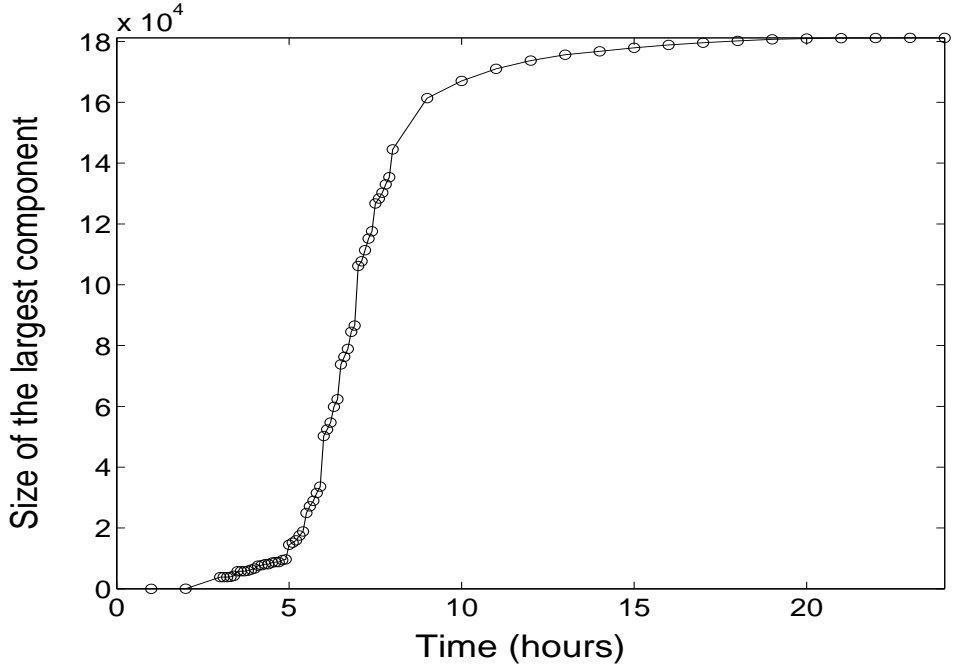}}
  \end{center}
 \caption{The size of the largest component (cluster) over time. A
   sharp transition is observed at about 6 a.m when people move from
   home to work or school.}
\label{myfig01}
\end{figure}

\begin{figure}[h*]
  \begin{center}
   \resizebox{6in}{3.5in}{\includegraphics{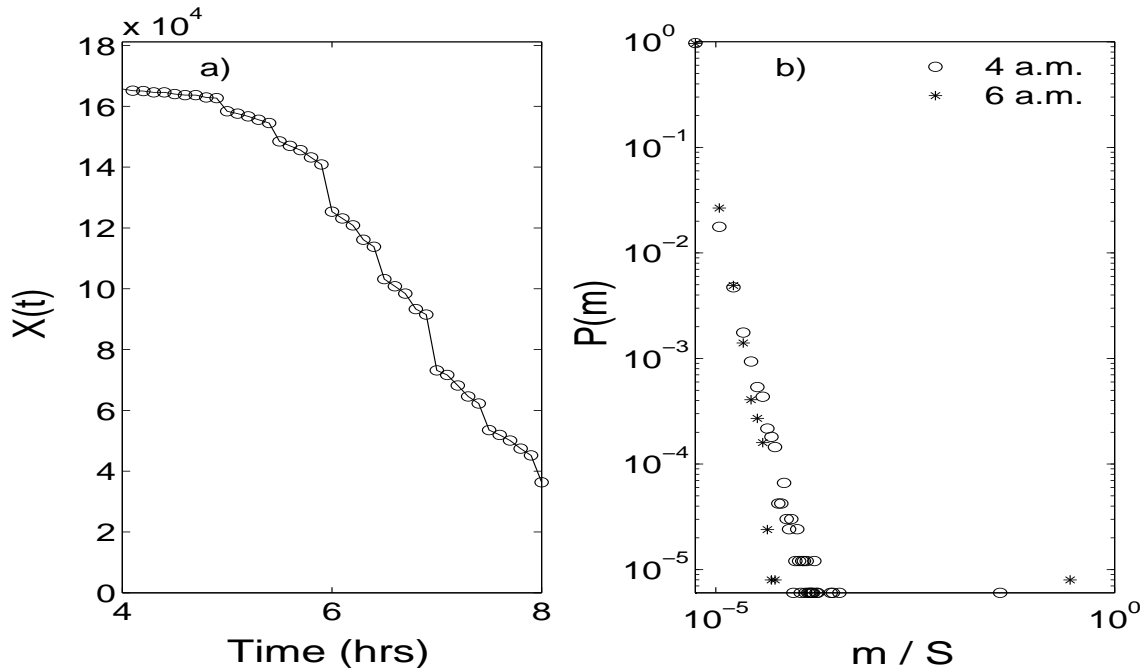}}
  \end{center}
 \caption{(a) The number of components $X(t)$ between $4$ a.m. and $8$
   a.m. (b) Probability distribution $P(m)$ of the normalized component sizes at two
   different times of the day. The component sizes ($m$) have been
   normalized by $S$, the size of the largest component of the cumulative
   network during the whole day (Table $1$).}
\label{myfig011}
\end{figure}

\begin{table}
\begin{center}
 \begin{tabular}{cc}
 Time (hrs) & Size of largest component\\
 \hline
 5.6 & 27,132 \\
 5.8 & 31,511 \\
 6.0 & 50,242 \\
 6.2 & 54,670 \\
 6.4 & 62,346 \\
 6.6 & 76,290 \\
 6.8 & 84,516 \\
 7.0 & 106,160 \\
 \end{tabular}
\end{center}
\label{table3}
\caption{Size of the largest component just before and after 6 a.m.,
 the time at which a sharp transition occurs. At midnight, all but 14
 locations belong to the largest component (Table $2$).}
\end{table}

Understanding the temporal properties of networks is critical to the study
of superimposed dynamics such as the spread of epidemics on networks. Most studies of 
superimposed processes on networks assumes that the contact structure
is fixed (see for example \cite{Grassberger1, Newman4, Newman5,
Moore1, Pastor1, Pastor2, May1, Zoltan1, Eguiluz1}). Here, we take a
look at the time evolution of the largest connected component of the
location-based network of the city of Portland (Figure
\ref{myfig01}). We have observed that a sharp transition occurs at
about $6$ a.m. In fact, by $7$ a.m. the size of the largest component
includes approximately $60\%$ of the locations (nodes). Table $3$
shows the size of the largest component just before and after the sharp 
transition occurs.\\

\indent Let $X_m(t)$ be the number of components of size $m$
at time $t$. Then $X(t) = \sum_{m \ge 1} X_m(t)$ is the total number
of components at time t (Figure \ref{myfig011}(a)). Furthermore, the
probability $P(m)$ that a randomly chosen node (location) belongs to a
component of size $m$ follows a power law that gets steeper in time as
the giant component forms (Figure \ref{myfig011}(b)).\\

To identify the relevance of the temporal trends, we computed the out-degree 
distribution of the network for three different time intervals: The morning from 
$6$ a.m to $12$ p.m.; the workday from $6$ a.m. to 6 p.m.; and the
full $24$ hours. In the morning phase, the 
out-degree distribution has a tail that decays as a power law with 
$\gamma \simeq 2.7$ (for the workday $\gamma \simeq 2.43$ and for the full day 
$\gamma \simeq 2.4$). The distribution of the out-degree data has two 
scaling regions: the number of locations is approximately constant for 
out-degree $k<20$ and then decays as a power law for 
high degree nodes (Fig. \ref{myfig1}). The degree distribution for the
undirected network (ignoring edge direction) displays power-law behavior,
but with slightly different power-law exponents: $2.3$ (morning), $2.48$
(work day) and $2.51$ (full day).

\begin{figure}[h*]
  \begin{center}
    \resizebox{6in}{3in}{\includegraphics{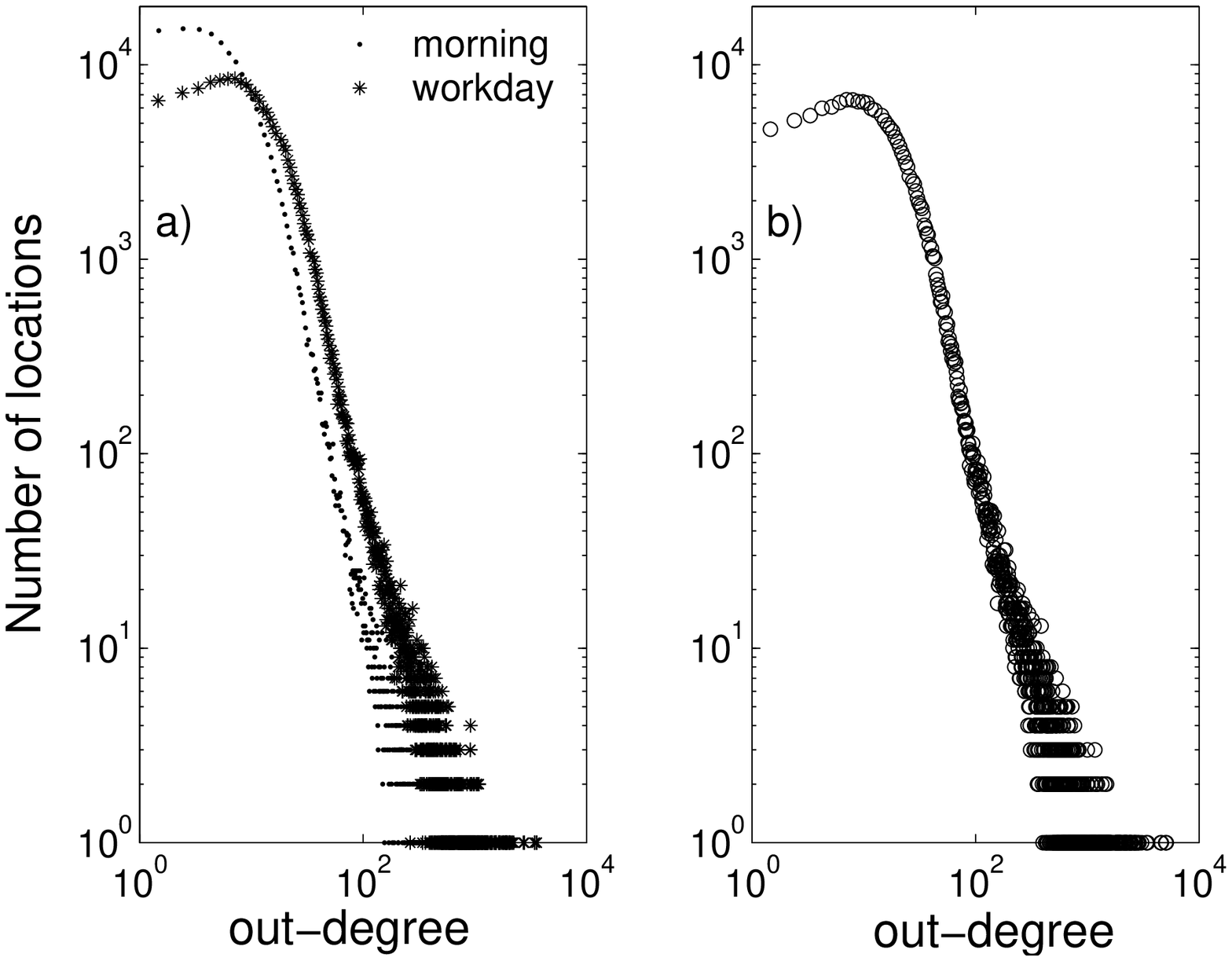}}
  \end{center}
 \caption{Distribution of the out-degrees of the location-based
   network of the city of Portland. There are approximately 
the same number of nodes (locations) with out-degree $k=1,2, ...10$. For $k>10$ the 
number of nodes with a given out-degree decays as a power law $P(k) 
\propto k^{-\gamma}$ with (a) $\gamma \simeq 2.7$ for the morning (6 a.m.-12 
p.m.), $\gamma \simeq 2.43$ for the workday (6 a.m.-6 p.m.) and (b) $\gamma \simeq 2.4$ 
for the full day.}
\label{myfig1}
\end {figure}

\begin{figure}[h*]
  \begin{center}
    \resizebox{6in}{3in}{\includegraphics{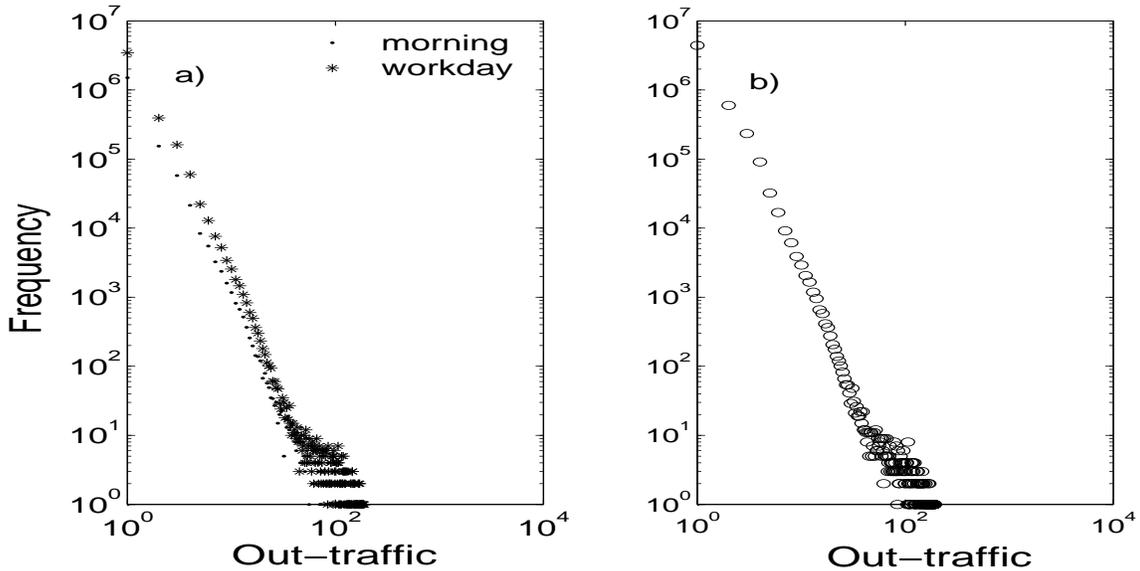}}
  \end{center}
 \caption{The out-traffic distribution of the location-based network of the 
city of Portland follows a power law ($P(k) \propto k^{-\gamma}$)
with (a) $\gamma \approx 3.56$ (morning), $\gamma \approx 3.74$
(afternoon), and (b) $\gamma \approx 3.76$ (full day). Hence a few
connections have high traffic but most connections have
low traffic.}
\label{myfig2}
\end {figure}

\begin{figure}[h*]
  \begin{center}
    \resizebox{5in}{3.5in}{\includegraphics{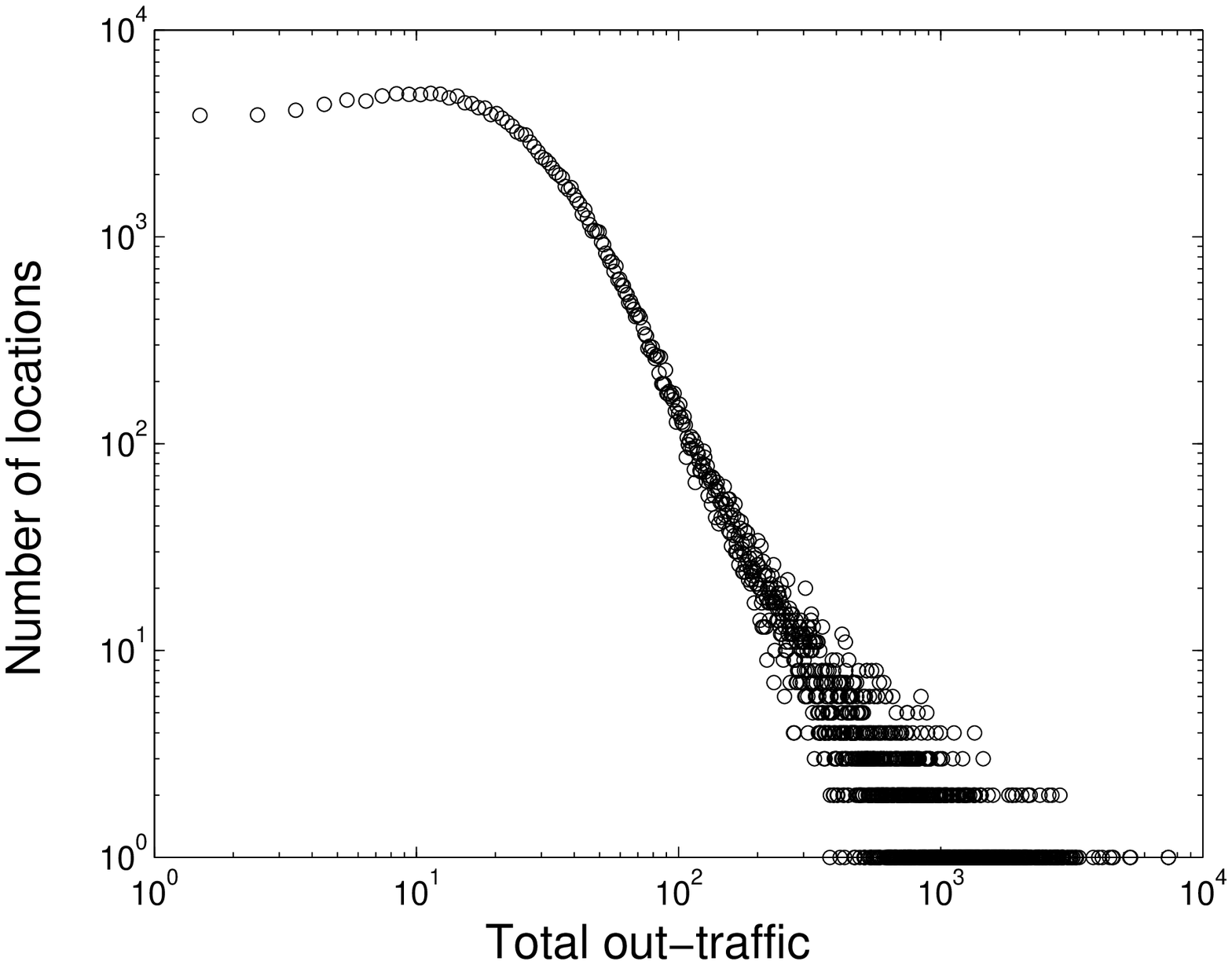}}
  \end{center}
 \caption{Distribution of the total out-traffic for the location-based 
network of the city of Portland. There are approximately 
the same number of locations (nodes) with small total out-traffic. 
The number of locations where more than 30 people (approximately) leave each 
day decays as a power law with $\gamma \simeq 2.74$.}
\label{myfig3}
\end {figure}

The strength of the connections in the location-based network is 
measured by the traffic (flow of individuals) between locations in a 
``typical'' day of the city of Portland. The log-log plot of the out-traffic 
distributions for three different periods of time (Fig. \ref{myfig2}) exhibits power law 
decay with exponents, $\gamma \simeq 3.56$ for the morning, $\gamma \simeq 3.74$ for the 
workday, and $\gamma \simeq 3.76$ for the full day. The 
out-traffic distribution is characterized by a power law 
distribution for all values of the traffic-weight matrix $W$.  This is not 
the case for the out-degree distribution of the network (see Figure 
\ref{myfig1}) where a power law fits well only for sufficiently large 
degree $k$ ($k>10)$. \\
\indent The distribution of the total out-traffic per location, 
$w_i$'s ($w_i = \sum_j w_{i,j}$), is characterized by two scaling
regions. The tail of this distribution decays as a power
law with exponent $\gamma = 2.74$ (Fig. \ref{myfig3}). This is almost
the same decay as the out-degree distribution ($\gamma = 2.7$) because
the out-degree and the total out-traffic are highly correlated (with
correlation coefficient $\rho = 0.94$).

\section{Correlation between out-degree and total out-traffic}

\begin{figure}[h*]
  \begin{center}
    \resizebox{4.5in}{3.5in}{\includegraphics{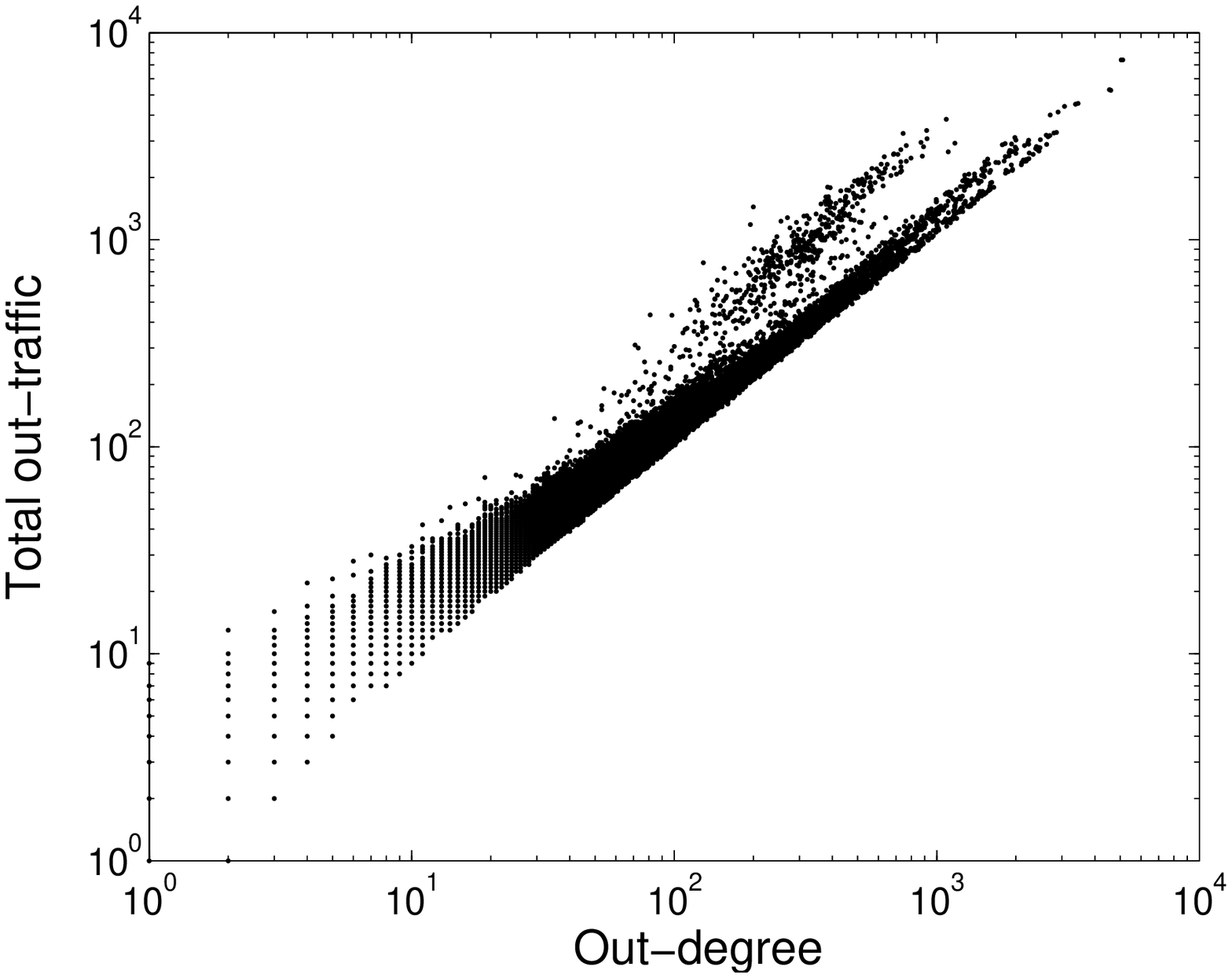}}
  \end{center}
 \caption{ Correlation between the out-degree and the total
   out-traffic. The correlation coefficient is $\rho=0.94$ on a
   log-log scale. Most ($95 \%$) of the locations have fewer 
   than $100$ people leaving during the day.}
\label{myfig4}
\end {figure}

The degree of correlation between various network 
properties depend on the social dynamics of the population. The
systematic generation and resulting structure of these networks is
important to understand dynamic processes such as epidemics that
``move'' on these networks. Understanding the mechanisms behind these
correlations will be useful in modeling fidelity networks.\\

\indent In the Portland network, the out-degree $k$ and total out-traffic $v$ have a correlation 
coefficient $\rho=0.94$ on a log-log scale with $95\%$ of the nodes (locations) having out-degree and total 
out-traffic less than $100$ (Fig. \ref{myfig4}). That is, the density of their joint 
distribution $F(k,v)$ is highly concentrated near small values of the out-degree 
and total out-traffic distributions. The joint distribution supports a surface 
that decays linearly when the density is in $log_e$ scale (Figure \ref{myfig5}). 


\section{Conclusions}

Strikingly similar patterns on data from the movement of $1.6$ million individuals in a ``typical'' day in 
the city of Portland have been identified at multiple temporal scales and various levels of aggregation.
The analysis is based on the mapping of people's movement on a weighted directed graph
where nodes correspond to physical locations and where directed edges, connecting the nodes, are weighted
by the number of people moving in and out of the locations during a 
typical day. The clustering coefficient, measuring the
local connectedness of the graph, scales as $k^{-1}$ ($k$ is the degree 
of the node) for sufficiently large $k$. This scaling is consistent
with that obtained from models that postulate underlying hierarhical structures (few nodes get most of the 
action). The out-degree distribution in log-log scale is relatively 
constant for small $k$ but exhibits power law decay afterwards ($P(k) \propto 
k^{-\gamma}$). The distribution of daily total 
out-traffic between nodes in log-log scale is flat for small $k$ but 
exhibits power law decay afterwards.
The distribution of the daily out-traffic of individuals 
between nodes scales as a power law for all $k$ (degree).\\
\indent The observed power law distribution in the out-traffic (edge 
weights) is therefore, supportive of the theoretical
analysis of Yook \textit{et al.} \cite{Yook} who built weighted 
scale-free (WSF) dynamic networks and proved that 
the distribution of the total weight per node (total out-traffic in our 
network) is a power law where the weights are exponentially distributed.\\
\indent There have been limited attempts to identify at least some characteristics of the joint
distributions of network properties. The fact that daily out-degree
and total out-traffic data are highly correlated is consistent again
with the results obtained from models that assume an underlying hierarhical structure (few nodes have most of the connections and get most of 
the traffic (weight)). The Portland network
exhibits a strong linear correlation between out-degree and total 
out-traffic on a log-log scale. We use this time series data
to look at the network ``dynamics''. As the activity in the network
increases, the size of the maximal connected component exhibits
threshold behavior, that is, a ``giant'' connected component, suddenly
emerges. The study of superimposed processes on networks such as those
associated with the potential deliberate release of biological agents
needs to take into account the fact that traffic is not
constant. Planning, for example, for worst-case scenarios requires
knowledge of edge-traffic, in order to characterize the temporal
dynamics of the largest connected network components \cite{Chowell2}.

\section{Acknowledgements}

The authors thank Pieter Swart, Leon Arriola, and Albert-L\'{a}szl\'{o} 
Barab\'{a}si for interesting and helpful discussions. This research
was supported by the Department of Energy under contracts
W-7405-ENG-36 and the National Infrastructure Simulation and Analysis
Center (NISAC).


\begin{thebibliography}{1}
\bibitem{comment1} $m$ is constant for the BA model. We have used
  $m=16$, the median out-degree of our network.
\end{thebibliography}
\end{document}